\documentclass[12pt,a4paper]{article}  
\usepackage{psfrag}   
\usepackage{graphics}  
\usepackage{amssymb,epsfig,amsmath,euscript,array} 
\usepackage{cite}  
\usepackage{epsfig}
\usepackage{slashbox}
\usepackage{amsmath,amssymb}

\makeatletter
\@addtoreset{equation}{section}
\makeatother
\renewcommand{\theequation}{\thesection.\arabic{equation}}

\newcommand{\startappendix}{
\setcounter{section}{0}
\renewcommand{\thesection}{\Alph{section}}
\renewcommand{\theequation}{\Alph{section}.\arabic{equation}}}

\newcommand{\Appendix}[1]{
\refstepcounter{section}
\begin{flushleft}
{\Large\bf Appendix \thesection:
#1}
\end{flushleft}}

\newcounter{multieqs}

\newcommand{\be}{\begin{equation}}  
\newcommand{\ee}{\end{equation}}

\newcommand{\bm}[1]{\mbox{\boldmath $#1$}}

\def\hf{{\textstyle{1\over2}}}
\def\ihf{{\textstyle{i\over2}}}

\def\bd{\begin{document}}  
\def\ed{\end{document}}  
\def\nn{\nonumber}  
\def\bea{\begin{eqnarray}}  
\def\eea{\end{eqnarray}}  
\let\bm=\bibitem  
\let\la=\label  
  
\def\npb#1#2#3{Nucl. Phys. {\bf{B#1}} #3 (#2)}  
\def\plb#1#2#3{Phys. Lett. {\bf{#1B}} #3 (#2)}  
\def\prl#1#2#3{Phys. Rev. Lett. {\bf{#1}} #3 (#2)}  
\def\prd#1#2#3{Phys. Rev. {D \bf{#1}} #3 (#2)}  
\def\cmp#1#2#3{Comm. Math. Phys. {\bf{#1}} #3 (#2)}  
\def\cqg#1#2#3{Class. Quantum Grav. {\bf{#1}} #3 (#2)}  
\def\nppsa#1#2#3{Nucl. Phys. B (Proc. Suppl.) {\bf{#1A}}#3 (#2)}  
\def\ap#1#2#3{Ann. of Phys. {\bf{#1}} #3 (#2)}  
\def\ijmp#1#2#3{Int. J. Mod. Phys. {\bf{A#1}} #3 (#2)}  
\def\rmp#1#2#3{Rev. Mod. Phys. {\bf{#1}} #3 (#2)}  
\def\mpla#1#2#3{Mod. Phys. Lett. {\bf A#1} #3 (#2)}  
\def\jhep#1#2#3{J. High Energy Phys. {\bf #1} #3 (#2)}  
\def\atmp#1#2#3{Adv. Theor. Math. Phys. {\bf #1} #3 (#2)}  
  
\newcommand{\EQ}[1]{\begin{equation} #1 \end{equation}}  
\newcommand{\AL}[1]{\begin{subequations}\begin{align} #1 \end{align}\end{subequations}}  
\newcommand{\SP}[1]{\begin{equation}\begin{split} #1 \end{split}\end{equation}}  
\newcommand{\ALAT}[2]{\begin{subequations}\begin{alignat}{#1} #2 \end{alignat}\end{subequations}}  
\def\beqa{\begin{eqnarray}}   
\def\eeqa{\end{eqnarray}}   
\def\beq{\begin{equation}}   
\def\eeq{\end{equation}}   
  
\def\N{{\cal N}}  
\def\sst{\scriptscriptstyle}  
\def\thetabar{\bar\theta}  
\def\Tr{{\rm Tr}}  
\def\one{\mbox{1 \kern-.59em {\rm l}}}  
 \def\Nh{\hat{N}} 
  
\def\a{\alpha}      \def\da{{\dot\alpha}}  
\def\b{\beta}       \def\db{{\dot\beta}}  
\def\c{\gamma}  \def\G{\Gamma}  \def\cdt{\dot\gamma}  
\def\d{\delta}  \def\D{\Delta}  \def\ddt{\dot\delta}  
\def\e{\epsilon}        \def\vare{\varepsilon}  
\def\f{\phi}    \def\F{\Phi}    \def\vvf{\f}  
\def\h{\eta}  
\def\k{\kappa}  
\def\l{\lambda} \def\L{\Lambda}  
\def\m{\mu} \def\n{\nu}  
\def\o{\omega}  
\def\p{\pi} \def\P{\Pi}  
\def\r{\rho}  
\def\s{\sigma}  \def\S{\Sigma}  
\def\t{\tau}  
\def\th{\theta} \def\Th{\Theta} \def\vth{\vartheta}  
\def\X{\Xeta}  
\def\z{\zeta}  
  
\def\cA{{\cal A}} \def\cB{{\cal B}} \def\cC{{\cal C}}  
\def\cD{{\cal D}} \def\cE{{\cal E}} \def\cF{{\cal F}}  
\def\cG{{\cal G}} \def\cH{{\cal H}} \def\cI{{\cal I}}  
\def\cJ{{\cal J}} \def\cK{{\cal K}} \def\cL{{\cal L}}  
\def\cM{{\cal M}} \def\cN{{\cal N}} \def\cO{{\cal O}}  
\def\cP{{\cal P}} \def\cQ{{\cal Q}} \def\cR{{\cal R}}  
\def\cS{{\cal S}} \def\cT{{\cal T}} \def\cU{{\cal U}}  
\def\cV{{\cal V}} \def\cW{{\cal W}} \def\cX{{\cal X}}  
\def\cY{{\cal Y}} \def\cZ{{\cal Z}}

\def\ua{\underline{\alpha}}  
\def\ub{\underline{\phantom{\alpha}}\!\!\!\beta}  
\def\uc{\underline{\phantom{\alpha}}\!\!\!\gamma}  
\def\um{\underline{\mu}}  
\def\ud{\underline\delta}  
\def\ue{\underline\epsilon}  
\def\una{\underline a}\def\unA{\underline A}  
\def\unb{\underline b}\def\unB{\underline B}  
\def\unc{\underline c}\def\unC{\underline C}  
\def\und{\underline d}\def\unD{\underline D}  
\def\une{\underline e}\def\unE{\underline E}  
\def\unf{\underline{\phantom{e}}\!\!\!\! f}\def\unF{\underline F}  
\def\unm{\underline m}\def\unM{\underline M}  
\def\unn{\underline n}\def\unN{\underline N}  
\def\unp{\underline{\phantom{a}}\!\!\! p}\def\unP{\underline P}  
\def\unq{\underline{\phantom{a}}\!\!\! q}  
\def\unQ{\underline{\phantom{A}}\!\!\!\! Q}  
\def\unH{\underline{H}}  
  
\def\As {{A \hspace{-6.4pt} \slash}\;}  
\def\bs {{b \hspace{-6.4pt} \slash}\;}  
\def\Ds {{D \hspace{-6.4pt} \slash}\;}  
\def\ds {{\del \hspace{-6.4pt} \slash}\;}  
\def\ss {{\s \hspace{-6.4pt} \slash}\;}  
\def\ks {{ k \hspace{-6.4pt} \slash}\;}  
\def\ps {{p \hspace{-6.4pt} \slash}\;}  
\def\pas {{{p_1} \hspace{-6.4pt} \slash}\;}  
\def\pbs {{{p_2} \hspace{-6.4pt} \slash}\;}  
  
\def\Fh{\hat{F}}  
\def\Vh{\hat{V}}  
\def\Xh{\hat{X}}  
\def\ah{\hat{a}}  
\def\xh{\hat{x}}  
\def\yh{\hat{y}}  
\def\ph{\hat{p}}  
\def\xih{\hat{\xi}}  
  
\def\psit{\tilde{\psi}}  
\def\Psit{\tilde{\Psi}}  
\def\tht{\tilde{\th}}  
   
\def\At{\tilde{A}}  
\def\Qt{\tilde{Q}}  
\def\Rt{\tilde{R}}  
\def\Nt{\tilde{N}}  
  
\def\at{\tilde{a}}  
\def\st{\tilde{s}}  
\def\ft{\tilde{f}}  
\def\pt{\tilde{p}}  
\def\qt{\tilde{q}}  
\def\vt{\tilde{v}}  
\def\nt{\tilde{n}}  
  
\def\delb{\bar{\partial}}  
\def\bz{\bar{z}}  
\def\bD{\bar{D}}  
\def\bB{\bar{B}}  
  
\def\bk{{\bf k}}  
\def\bl{{\bf l}}  
\def\bp{{\bf p}}  
\def\bq{{\bf q}}  
\def\br{{\bf r}}  
\def\bx{{\bf x}}  
\def\by{{\bf y}}  
\def\bR{{\bf R}}  
\def\bV{{\bf V}}  
  
\def\d{\delta}\def\D{\Delta}\def\ddt{\dot\delta}  
  
\def\pa{\partial} \def\del{\partial}  
\def\xx{\times}  
\def\uno{\mbox{1 \kern-.59em {\rm l}}}    
  
\def\trp{^{\top}}  
\def\inv{^{-1}}  
\def\dag{{^{\dagger}}}  
\def\pr{^{\prime}}  
  
\def\rar{\rightarrow}  
\def\lar{\leftarrow}  
\def\lrar{\leftrightarrow}  
  
\newcommand{\0}{\,\!}      
\def\one{1\!\!1\,\,}  
\def\im{\imath}  
\def\jm{\jmath}  
  
\newcommand{\tr}{\mbox{tr}}  
\newcommand{\slsh}[1]{/ \!\!\!\! #1}  
  
\def\vac{|0\rangle}  
\def\lvac{\langle 0|}  
  
\def\hlf{\frac{1}{2}}  
\def\ove#1{\frac{1}{#1}}  

\def\Box{\square}  
\def\ZZ{\mathbb{Z}}  
\def\CC#1{({\bf #1})}  
\def\bcomment#1{}  
\def\bfhat#1{{\bf \hat{#1}}}  
\def\VEV#1{\left\langle #1\right\rangle}  

\newcommand{\ex}[1]{{\rm e}^{#1}} \def\ii{{\rm i}}  

\def\rr{{\rm r}} \def\rs{{\rm s}}\def\rv{{\rm v}}
\def\ri{{\rm i}}\def\rj{{\rm j}}
  
\newcommand{\lrbrk}[1]{\left(#1\right)}
\newcommand{\sfrac}[2]{{\textstyle\frac{#1}{#2}}}

\font\mybb=msbm10 at 12pt
\def\bb#1{\hbox{\mybb#1}}

\font\myBB=msbm10 at 18pt
\def\BB#1{\hbox{\myBB#1}}

\begin{document}  
  
\hfill{ hep-th/0308008}  
   
\vspace{20pt}  
   
\begin{center}

{\Large \bf  Effective actions,  Wilson lines and the IR/UV mixing   \\}
\vspace{10pt}
{\Large \bf  in   noncommutative  supersymmetric  gauge  theories}

\vspace{30pt}  
   
{\bf Jonathan Levell and  Gabriele Travaglini$^\sharp$  }

{\small \em
Centre for Particle Theory,
Department of Physics and IPPP,\\
University of Durham, Durham, DH1 3LE, UK
}

\vspace{10pt}  
  
{\sffamily \tt 
jonathan.levell@durham.ac.uk,
g.travaglini@qmul.ac.uk }

\vspace{30pt}  
{\bf Abstract}  
  
\end{center}  
We study IR/UV mixing effects in  noncommutative supersymmetric
Yang-Mills theories with gauge group $U(N)$ using 
background field perturbation theory.
We compute three- and four-point functions of background fields, 
and show that  the IR/UV mixed contributions to these correlators
can be reproduced from an explicitly gauge-invariant 
effective action, which is expressed in terms of 
open Wilson lines.

\vspace{7cm}  
\noindent
{\small $\sharp$ Present address: 
Department of Physics, Queen Mary College, London E1 4NS}
  
\setcounter{page}{0}  
\thispagestyle{empty}  
\newpage

\section{Introduction}
Quantum field theories on noncommutative spaces have 
attracted a lot of attention in the last few years. 
Part of this interest is motivated by 
the fact that noncommutative gauge theories appear
as the low-energy limit of open strings 
in the presence of a constant $B$ field \cite{CDS,DH,SW,CH}. 
Noncommutative theories are also very interesting from 
the purely field theoretical perspective, 
as they manifest novel features compared to 
their commutative cousins. 
The most interesting one is probably 
infrared/ultraviolet (IR/UV) mixing 
\cite{Minwalla,MST}.%
\footnote{For recent reviews of noncommutative 
theories and their relation to string theory, 
see \cite{dn,szabo}.}
Contrary to naive expectations based on commutative intuition,
the high-energy degrees of freedom of noncommutative theories 
in general  affect the physics at low-energy. 
In perturbation theory, 
the loop integrals in the planar sector 
of the noncommutative theory are exactly the same as in the commutative
counterpart, which implies the same structure of divergences and 
same counterterms of the commutative theory.
However, nonplanar diagrams are multiplied by phase
factors of the form $e^{ik\cdot\theta\cdot p}$,
where $k$ are external momenta and $p$ are loop momenta. 
This phase factor   improves the UV-convergence of nonplanar diagrams, 
and  typically renders  them finite.
But in  the IR limit of the external momenta,  
$  e^{\, ik\cdot\theta\cdot p}\rightarrow 1$
as $k\rightarrow 0 $, 
the nonplanar diagrams become divergent and this is now 
an IR-divergence.
%
This dramatically invalidates the naive expectations about a 
universal behaviour in the infrared for commutative 
and noncommutative theories, and  
the infrared regime of a noncommutative theory is 
in general  different from that 
of its commutative counterpart
\cite{Minwalla,MST}.

To explore this point further, 
in \cite{KT,HKT}  the low-energy wilsonian 
effective action  for a large class of supersymmetric 
and non-supersymmetric theories was  computed.   
The result was that, in the low-energy effective action,
the $U(1)$ degrees of freedom decouple from the $SU(N)$ part, with 
IR/UV mixing affecting only the $U(1)$ part 
of the gauge group \cite{HKT}. 
The leading order terms in the derivative expansion of the 
wilsonian effective action read 
\cite{KT,HKT}:
\be
\label{comm_action}
{S}_{\rm eff}  \ \ni \ 
{1  \over 4g^{2}_{\sst 1}(k) } 
 \   \int \! d^4x\, F^{\sst U(1)}_{\mu \nu}   
F^{\sst U(1)}_{\mu \nu}    
\ + \ {1  \over 4g^{2}_{\sst N}(k) } 
 \   \int \! d^4x\,F^{\sst SU(N)}_{\mu \nu}   
F^{\sst SU(N)}_{\mu \nu} 
\ , 
\ee
where the  coefficients in front of the gauge kinetic terms in
\eqref{comm_action}
define the wilsonian coupling constants 
of the corresponding gauge factors. 
The running of the $U(1)$ 
has the following asymptotic behaviour \cite{KT,HKT}:
\be
{1\over g^{2}_{\sst 1}(k)}  \, \rightarrow \, 
\pm {
\a_0 \over (4\pi)^2} \log k^2 \ , 
\label{r1}
\ee
where the plus (minus) sign corresponds to $k^2\to\infty$ 
($k^2\to 0$), whereas for the $SU(N)$ gauge factor we have,
in both limits, the usual (UV asymptotically free) running:
\be
{1\over g^{2}_{\sst N}(k)}  \, \rightarrow \, 
{
\b_0 
\over (4\pi)^2} \log k^2 
\ .
\label{rn}
\ee
In the previous expressions \eqref{r1}, \eqref{rn}, 
$\b_0$ is the first coefficient of the microscopic 
$\beta$-function, 
and $\a_0$ is a   positive constant whose 
numerical value 
is  determined by the field content of the theory, 
e.g.~$\a_0= \b_0 \equiv 3N $ 
for pure $U(N)$ $\cN=1$ super Yang-Mills
(for more general  situations see 
\eqref{alfazero} and \cite{KT,HKT,CKT1}).
The change in the running of the $U(1)$ coupling in \eqref{r1}
is the manifestation  of the IR/UV mixing, and 
occurs at a scale
$k^2 \sim M_{\sst \rm NC}^2$,  
where $M_{\sst \rm NC}\sim \theta^{-1/2}$ 
is the noncommutative mass.%
\footnote{We summarise our notation and  conventions 
in the  Appendix.}

In  any non-supersymmetric theory, 
quadratic divergences appear in the gauge field 
polarisation tensor
which would change the photon dispersion 
relation \cite{MST}, 
and hence threaten the renormalizability of the theory.
Interestingly, 
these quadratic divergences cancel in any supersymmetric theory, 
and in these theories we are left  
with the logarithmic infrared divergences 
in the effective $U(1)$ coupling encoded in 
\eqref{r1} \cite{MST,KT}.
Hence, supersymmetry appears to be a necessary 
ingredient for a noncommutative theory 
to be consistent  \cite{Minwalla,MST,KT}. 
For this reason, 
from now on we  will fix our attention on 
supersymmetric theories.

The peculiar behaviour discussed above 
for the $U(1)$ effective coupling constant 
was interpreted in \cite{CKT1} 
as having a full noncommutative $U(N)$ gauge theory in the 
ultraviolet, 
which in the low-energy limit appears as a commutative $SU(N)$ theory, 
with the $U(1)$ degrees of freedom which become
progressively more weakly coupled (i.e.~unobservable) in the infrared.
In the same paper, a mechanism for supersymmetry breaking 
was suggested where the $U(1)$ degrees of freedom 
act as the hidden sector, breaking supersymmetry at a scale 
potentially much lower than the noncommutativity scale,
and eventually 
becoming unobservable in the infrared  
due to the IR/UV mixing.
This IR/UV mixing  
therefore acquires the status of a very welcome feature of 
noncommutative theory, rather than being 
a field-theoretical illness of it.

The expression \eqref{comm_action} for the effective action 
would suggest that the noncommutative $U(N)$ gauge symmetry 
is broken at low energy; despite appearances, 
this is not the case. 
In \cite{Pernici:2000va} it was argued that the full one-loop
effective action for the $\cN=4$ theory is gauge invariant
(see also
\cite{Zanon,Santambrogio:2000rs,Liu:2000ad});
nonplanar diagrams give gauge-noninvariant 
contributions to e.g.~the {\it four}-point function, but 
in the Ward identities these terms are 
precisely cancelled by  
gauge-noninvariant terms in the 
{\it five}-point function \cite{Pernici:2000va}.
This mechanism of cancellations between 
different $n$-point functions is quite 
a clear clue for the presence of Wilson lines 
in the expression for the effective action. 
Indeed, in  
\cite{AL,VanRaamsdonk:2001jd},  
a gauge-invariant completion
of  \eqref{comm_action}
was proposed  which involves open Wilson lines
\cite{Das,Gross:observables,IIKK}.
It was conjectured  in \cite{AL} that the 
the IR/UV mixed contribution 
to the effective action of a supersymmetric gauge theory
can be reproduced by the following $U(N)$ gauge-invariant 
term:
\begin{equation}
\label{AL-conj}
S^{(1)}_{\rm eff}\, =\, 
-
\frac{\cC}{4} \,  \int\!\!  \frac{d^4 p}{(2
\pi)^4} \, 
{\cal O}^{\mu \nu}(-p) \, 
T(p)\, 
{\cal O}_{\mu \nu} (p)
\ , 
\end{equation}
where 
the gauge-invariant
operator ${\cal O}_{\mu \nu}$ is defined by
\begin{equation}
\label{O_def} 
{\cal O}_{\mu \nu} (p) = \Tr \int\! d^4 x\ L_\star
\left( F_{\mu \nu} (x) \, e^{
\, 
\int_0^1 \!\! d \sigma \, {\tilde
p}^\mu A_{\mu}(x+{\tilde p}\, \sigma)}\right) \star e^{i p x} 
\ ,
\end{equation}
and $L_{\star}$ stands for 
integration along the open Wilson line 
together with path ordering with respect to the star-product 
\cite{Liu:2000mj}. 
The matching of \eqref{AL-conj} with 
the analytic results 
\cite{KT} for the $U(1)$ effective coupling constant 
$1 / g^2_1 (k)$ of \eqref{r1}
determines 
the function $T(p)$  
and the numerical constant 
$\cC$ appearing in \eqref{AL-conj}. 
One easily finds  that  
\begin{equation}
\label{ti}
T(p)= \frac{2}{(4\pi)^2}
\int_{0}^{1} \! \! dx \, K_{0} (\sqrt{x(1-x)}\,
|p| | \tilde{p} |)
\ , 
\end{equation}
where $ K_{0}$ is a Bessel function, 
with $ K_0 (z) \to -\log (z/2)$ as $z\to 0$.
Moreover, $\cC =2 \a_0 / N$, where 
$\a_0$ is defined in \eqref{r1}.

In this paper we would like confirm the interesting conjecture  
of \cite{AL} with an explicit field theory calculation.
To start probing the presence of the Wilson line operator in
\eqref{O_def} we need to calculate an $n$-point function 
of gauge fields with $n\geq 3$; 
for this reason, we will concentrate on 
the cases of three- and four-point functions
of background fields.
Our formalism is, however, general, and allows 
in principle 
to calculate generic $n$-point correlators.
We will focus our attention on a 
generic $\cN =1$ supersymmetric field theory 
with $N_{\rm f}$ adjoint chiral multiplets%
\footnote{We would like to remind the reader that 
the vacuum polarisation tensor, and therefore the 
wilsonian coupling constant, receive 
nonplanar (i.e.~IR/UV mixed) contributions 
only from fields in the adjoint representation. 
Fields in the fundamental representation 
do not contribute to the IR/UV mixing
\cite{KT}, and are therefore irrelevant
for our analysis.
This circumstance was first noticed 
in noncommutative QED in \cite{haya}.} 
and make use of the background field method.
The case $N_{\rm f}=0 $ corresponds 
to pure $\cN=1$ super Yang-Mills, 
whereas for $N_{\rm f}=1, 3 $ we have the $\cN=2$ and $\N=4$ 
theories, respectively.
The results of our computations  
confirm the presence of the term 
\eqref{AL-conj} in the effective action. However, 
our results also show that we need  
to include another term $S^{(2)}_{\rm eff}$ 
in the effective action, 
which can be written as
%
\begin{equation}
S^{(2)}_{\rm eff} = 
\frac{{\cC}}{2}
\int\!\!  \frac{d^4 p }{(2 \pi)^4} 
\, 
 {\cal O}_{F^2} (p) \ 
W'( - p) 
\, T(p) 
\ , 
\label{new_new_term}
\end{equation}
where
\begin{equation}
\label{OFsquared}
{\cal O}_{F^2} (p)={\Tr } \int\! d^4 x \ 
L_\star \left( F_{\mu \nu}(x) 
F^{\mu \nu}(x) e^{
\, \int_0^1 \!\! d \sigma \, \tilde{p}^\mu
A_{\mu}(x+\tilde{p} \sigma)}
\star e^{ipx} \right)
\, \ , 
\end{equation}
and ${\cal W'} (p)$ denotes the open Wilson line operator
${\cal W} (p)$ with the ${\cal O} (A^0 )$ term subtracted, 
where,  in our conventions 
\begin{equation}
\label{W} 
{\cal W} (p) = \Tr \int\! d^4 x\ P_\star
\left( e^{
\, 
\int_0^1 \!\! d \sigma \, {\tilde
p}^\mu A_{\mu}(x+{\tilde p}\, \sigma)}\right) \star e^{i p x} 
\ . 
\end{equation}
$S^{(2)}_{\rm eff}$ is 
manifestly gauge invariant, 
and again contains open Wilson lines. 

The appearance of the term $S^{(2)}_{\rm eff}$
in   \eqref{new_new_term} is not  unexpected.  
Indeed, 
similar contributions were  predicted in 
\cite{Jiang:2001qa}, 
where a wilsonian calculation of 
the effective action was performed using
the matrix model approach to  
noncommutative gauge theories. 
Similar results were also obtained in 
\cite{Kiem:2001dm} using the bosonic world-line approach.%
\footnote{We thank Adi Armoni and an anonymous Referee for 
pointing the papers \cite{Jiang:2001qa,Kiem:2001dm} 
to our attention.}
In the first version of this paper 
we proposed to use the following expression, 
instead than \eqref{new_new_term}:
\begin{equation}
{\tilde{S}}^{(2)}_{\rm eff} = 
-
i\, \frac{\cC}{4}
\int\!\!  \frac{d^4 q_1}{(2 \pi)^4} 
\frac{d^4 q_2}{(2 \pi)^4}\, 
 {\cal Q}(-q_1 , -q_2) \ 
\theta^{\mu \nu} {\cal O}_{\mu \nu}(q_1+q_2) 
\, T(q_1 +q_2) 
\ , 
\label{new_term}
\end{equation}
where
\begin{equation}
{\cal Q}(p,q)={\Tr } \int\! d^4 x \ 
L_\star \left( F_{\mu \nu}(x) 
F^{\mu \nu}(x) e^{
\, \int_0^1 \!\! d \sigma \, \tilde{p}^\mu
A_{\mu}(x+\tilde{p} \sigma)}
\star e^{ipx} \right)
\, {\sin (q \tilde{p} / 2 ) \over q \tilde{p}}
\ .
\end{equation}
The interaction term in \eqref{new_term}, 
though   manifestly gauge invariant, 
is not satisfactory as it stands, 
since  $\theta^{\mu \nu}$ appears 
in \eqref{new_term} 
not only inside the Wilson lines but also explicitly.
This, in turns, would render its interpretation from 
the D-brane perspective  very difficult. 
It is easy to check that, once we expand 
the Wilson line in \eqref{new_new_term} up to  
${\cal O}({A})$, this term contributes 
to the three-point function in the same way as 
the term in \eqref{new_term} does. 
Indeed, we will show that \eqref{new_new_term}
and \eqref{new_term} both produce a contribution 
to the three-point function which is in precise
agreement with 
the direct calculation in the microscopic theory. 
Of course, at the level of four-point functions
\eqref{new_new_term} and 
\eqref{new_term} start producing 
contributions which are different. 
By comparing the perturbative result 
for a {\it four}-point function to the corresponding result 
derived from the effective action,  we will be able 
in the next sections to confirm that 
\eqref{new_new_term} is the correct expression  
to be incorporated in the 
effective action (rather than \eqref{new_term}), 
as also suggested by D-brane physics \cite{Jiang:2001qa}.

Let us mention that it would be very nice to have 
complete control on generic $n$-point functions
(or at least on the IR/UV mixed contributions), 
and use this knowledge to derive the full low--energy 
effective action. 
An interesting and  simple  way to evaluate the 
IR/UV mixed quadratically divergent  contributions (the poles)  
of generic $n$-point functions in a $U(1)$ non-supersymmetric 
gauge theory was devised in section \mbox{(3.1)} of
\cite{AL}. Unfortunately, simplifications similar 
to those exploited in \cite{AL} 
seem to be lacking in the case at hand.%
\footnote{The quadratic divergences in the effective action 
generated by IR/UV mixing 
are not present   for the case of  
{\it supersymmetric} gauge theories, where we are left only 
with logarithmic divergences. These are  more difficult 
to extract from the Feynman diagrams expressions.}
It would be interesting to 
apply the bosonic worldline approach, used 
in \cite{Kiem:2001dm} for non-supersymmetric 
theories,  to the case of supersymmetric theories
considered here, 
and see if that formalism  would lead to more tractable 
expressions than those obtained using 
conventional background perturbation theory.

The plan of the rest of this paper is as follows. 
In section 2 we will obtain  the contributions to the 
three- and four-point functions 
of gauge fields from  
the terms  $S^{(1)}_{\rm eff}$ and $S^{(2)}_{\rm eff}$
in the effective action, 
Eqs.~\eqref{AL-conj} and \eqref{new_new_term}, respectively.
Sections 3 contains the set-up for the application of the 
background field method to noncommutative gauge theories, 
and our Feynman rules.
Using the background field  method, we calculate in section 4 
the three- and four-point functions of background fields. 
In section 5 we compare the perturbative results 
derived in section 4 to the result obtained in section 2 from 
the effective action, finding agreement.
For other related work on noncommutative theories, 
see 
\cite{dn,szabo,Armoni2,MSR,Ter,GBM,Gross:dynamics,
Bonora,Zanon2,AMT,starwars,CKT2,CKT3,Banerjee:2001un,ALT,AGVM,ALU,
Banerjee:2003ce,Lopez,Ardalan:2003ev} 
and references therein.

Before concluding this introduction, 
we would like to mention that IR/UV mixing  also appears  
\cite{BFR} in  the recently discussed field theories on 
non(anti)commuting superspace \cite{Seiberg}.
Again, the resummed effective action appears to contain 
open Wilson lines in superspace \cite{BFR}.

\section{Three- and four-point functions from 
an effective action with  Wilson lines}
\subsection{The three-point function}
We begin by calculating the contribution 
from the effective interaction in 
\eqref{AL-conj} to the three-point function
\beq
\label{tpf}
\Gamma_{\m \n \r}^{ABC} (k_1 , k_2 , k_3 )
\,  := \,  
\int\!\!  \prod_{i=1}^{3} \, 
\frac{d^4 x_i}{(2 \pi)^4}\
e^{\, i \sum_{i=1}^{3} k_i  x_i } \, 
\langle \, 
A_{\m}^A (x_1)\,  A_{\n}^B (x_2)\,  A_{\r}^C (x_3)\,  
\rangle
\ . 
\eeq
In order to calculate the contribution from \eqref{AL-conj}
we need only to  expand the 
expression for $\cO_{\m \n}(p)$ in 
\eqref{O_def}
up to order $A^2$.
 We then Fourier transform and use \eqref{field-comm} and
\eqref{star-prod}, to get:
\beq
\label{mm}
{\cal O}_{\mu \nu} (p) = 
l_{\m \n}(p) + q^{(1)}_{\m \n}(p) + q^{(2)}_{\m \n}(p) + 
\cO (A^3)
\ , 
\eeq
where
\beqa
l_{\m \n}(p) &=& 
-\sqrt{N\over 2} 
\ 
\left[ p_\mu A_\nu^0 (p) - p_\nu A_\mu^0 (p)\right]
\ , 
\\ \nonumber
q^{(1)}_{\m \n}(p) &=&
i
A_\mu^A(q) A_\nu^A(p-q) \ \sin \frac{q \tilde{p}}{2}
\ , 
\\ \nonumber
q^{(2)}_{\m \n}(p)&=& i
\int\!\!  \frac{d^4 q}{(2 \pi)^4} \, \left[ q_{\mu}
A_{\nu}^A(q)  - q_{\nu} A_{\mu}^A(q)  \right]
(\tilde{p} \cdot A^A(p-q))   \ 
{ \sin (q \tilde{p} /  2) 
\over q \tilde{p}}
\ . 
\eeqa
Using  \eqref{mm}, we get  
the following contribution to the three-point function 
from the effective action $S^{(1)}_{\rm eff}$ 
of  \eqref{AL-conj}: 
\beqa  
\label{a_contr} 
\left. 
\Gamma^{ABC}_{\mu \nu \rho}(k_1, k_2,k_3)\right|_{S^{
(1)}_{\rm eff}}
& = &   
i
\cC \sqrt{\frac{N}{2}} \, {\sin ( k_{2}
\tilde{k}_{3} / 2 ) \over 
k_2 \tilde{k}_3}
\
(2\pi)^4\,\delta^{(4)}(k_{1}+k_{2}+k_{3}) 
\\ \nonumber \cr 
&&\cdot
\, 
\Bigl\{
T(k_{1})\  \delta^{A0} \delta^{BC}
\left[
 \tilde{k}_{1 {\nu}} P^{13}_{\r \m}
 +
 \tilde{k}_{1{\rho}}P^{12}_{\n \m}
+ (k_2 \tilde{k}_3)\, Q^1_{\m, \n \r} 
\right] 
 \\ \nonumber 
&&\ +   
T(k_{2})\ \delta^{B0} \delta^{CA} 
\left[  \tilde{k}_{2{\mu}}P^{23}_{\r \n}
+
 \tilde{k}_{2{\rho}} P^{12}_{\n \m} + 
 (k_2 \tilde{k}_3)\, Q^2_{\n, \r \m} 
 \right]
\\ \nonumber 
&&
 \ +
T(k_{3})\ \delta^{C0} \delta^{AB} 
\left[  \tilde{k}_{3 {\mu}} P^{23}_{\r \n} + 
 \tilde{k}_{3{\nu}} P^{13}_{\r \m} + 
 (k_2 \tilde{k}_3)\, Q^3_{\r, \m \n} 
\right] \Bigr\}
\ , 
\eeqa
where we have defined 
\beqa
P^{ij}_{\m\n} \ :=\  \d_{\m \n} (k^i \cdot k^j) \,  - \, k^i_\m k^j_\n  \ , 
\qquad 
Q^{i}_{\r, \m \n} \ :=\  \d_{\r \m} \, k^{i}_\n \, - \, \d_{\r \n} \, k^i_\m 
\ .
\label{PQ}
\eeqa 
In a similar way we can calculate the contribution to 
the three-point function from the term 
$S^{(2)}_{\rm eff}$  in 
\eqref{new_new_term},  obtaining:
\beqa
\label{new_term_contrib}
&& 
\left.
\Gamma^{ABC}_{\mu \nu \rho}(k_1, k_2,k_3)
\right|_{S^{(2)}_{\rm eff}}
=     i \cC \sqrt{\frac{N}{2}} \ 
{ \sin (k_2 \tilde{k}_3  / 2) \over 
k_2 \tilde{k}_3} 
\
(2\pi)^4\,\delta^{(4)}(k_{1}+k_{2}+k_{3}) 
\\ \nonumber
&& 
\qquad \qquad 
\cdot\, \left\{   
\delta^{A0} \delta^{BC}\,  T(k_1)
\, \tilde{k}_{1 \mu} P_{\r \n}^{23}
\, + \, 
\delta^{B0} \delta^{AC}\,   T(k_2)
\, \tilde{k}_{2 \nu} P_{\m\r}^{31} 
\, + \,  
\delta^{C0} \delta^{AB}\, T(k_3)
\, \tilde{k}_{3 \r} P_{\n \m}^{12}
\right\}
\ . 
\eeqa
\subsection{The four-point function}
In this section we compute the contributions 
to the four-point function obtained from the effective action 
$S^{(1)}_{\rm eff}$ and
$S^{(2)}_{\rm eff}$ given in 
\eqref{AL-conj}, \eqref{new_new_term}, respectively. 
For the sake of simplicity we will restrict ourselves to 
the case of noncommutative $U(1)$ gauge group, and compute the 
four-point function
\beq
\label{fopf}
\Gamma_{\m \n \r \s} (k_1 , k_2 , k_3 , k_4 )
\,  := \,  
\int\!\!  \prod_{i=1}^{4} \, 
\frac{d^4 x_i}{(2 \pi)^4}\
e^{\, i \sum_{i=1}^{4} k_i  x_i } \, 
\langle \, 
A_{\m} (x_1)\,  A_{\n}  (x_2)\,  A_{\r} (x_3)\,  
A_{\s} (x_4)\,
\rangle
\ . 
\eeq
The result for 
$\Gamma_{\m \n \r \s} (k_1 , k_2 , k_3 , k_4 )$
is better expressed in terms of the quantities 
$J_2(k_1 , k_2)$ and $J_3 (k_1, k_2, k_3) $ 
introduced in \cite{Liu:2000mj} (see Appendix B of that paper), 
where
\beqa
\label{jfuncs2}
J_2 (k_1 , k_2) & = & 
{
\sin {k_1 \tilde{k}_2 \over  2} 
\over
{k_1 \tilde{k}_2 \over  2} 
} 
 \ , 
\\ \cr
J_3 (k_1, k_2, k_3) &=& 
{\sin {k_2 \tilde{k}_3 \over  2} \,  
\sin {k_1 (\tilde{k}_2 +\tilde{k}_3) \over     2}
\over { ( k_1 + k_2 )  \tilde{k}_3 \over  2} \, 
 {k_1  (\tilde{k}_2 + \tilde{k}_3) \over 2}}
\ + \ 
{{\sin {k_1 \tilde{k}_3 \over  2} \,  
\sin {k_2 (\tilde{k}_1 +\tilde{k}_3)   \over  2}}
\over {{ ( k_1 + k_2 )  \tilde{k}_3 \over  2} \, 
{k_2  (\tilde{k}_1 + \tilde{k}_3) \over  2}
}
}
\ . 
\label{jfuncs3}
\eeqa
Not surprisingly,  the functions $J_2 (k_1 , k_2)$ and 
$J_3 (k_1, k_2, k_3)$ \cite{Liu:2000mj} arise  
in the context of noncommutative effective action
for the one-loop $F^4$ term in  $\cN=4$ super Yang-Mills. 

In the same way as it was done for the 
three-point function, one finds that the contribution to 
the four-point function generated by the term 
\eqref{AL-conj} is given by the following expression: 
{\small
\beqa
\label{S1-4pt-piece}
&&
\left. 
\Gamma_{\m \n \r \s}(k_1, k_2,k_3,k_4)
\right|_{S^{
(1)}_{\rm eff}}
\, = \,  \, \, 
\!\!\!\! 
(2\pi)^4\,\delta^{(4)}(k_{1}+ \cdots + k_{4}) 
\, 
\Bigl\{ 
\,2 \, T(k_4) \, \cdot
\\ \nonumber \cr 
&&  
\!\!\!\! \cdot
\Bigl[
 \tilde{k}^{4}_{\r} Q^{4}_{\s , \n \m}
\sin  {k_1 \tilde{k}_2 \over  2} J_2 (k_3 , k_4 ) 
 +
 \tilde{k}^{4}_{\n} Q^{4}_{\s , \r \mu}
\sin {k_1 \tilde{k}_3 \over  2} J_2 (k_2 , k_4 )
+
\tilde{k}^{4}_{\mu} Q^{4}_{\s , \r \n}
\sin {k_2 \tilde{k}_3 \over  2} J_2 (k_1 , k_4 )
\Bigr]
\\ \nonumber \cr 
&&- T(k_4) \, J_3 (k_1 , k_2 , k_3 )\,  
\Bigl[
P^{41}_{\m \s} \, 
\tilde{k}_{\n}^{4} \, \tilde{k}_{\r}^{4} 
\, + \, 
P^{42}_{\n \s} \, 
\tilde{k}_{\mu}^{4} \, \tilde{k}_{\r}^{4} 
\, +\,  
P^{43}_{\r \s} 
\, \tilde{k}_{\mu}^{4} \, \tilde{k}_{\n}^{4} 
\Bigr]
\\ \nonumber \cr 
&& -T(k_1 + k_2) \Bigl[
\delta_{\mu \r} \delta_{\n \s} 
\sin {k_1 \tilde{k}_2 \over 2}   
\sin {k_3 \tilde{k}_4 \over 2}   
+ \, {1\over 2} P^{13}_{\r \mu} 
(\tilde{k}_1 + \tilde{k}_2)_{\n}
(\tilde{k}_3 + \tilde{k}_4)_{\s}
\, J_2 (k_1 , k_2 ) J_2 (k_3 , k_4 ) 
\\ \nonumber \cr 
&&
 + \, 
{1\over 2} Q^{3}_{\m , \n  \r} 
\sin {k_1 \tilde{k}_2 \over 2} 
(\tilde{k}_3 + \tilde{k}_4)_{\s}
\, J_2 (k_3 , k_4 ) \, 
+\,  {1\over 2} \, Q^{1}_{\r , \s  \mu} 
\sin {k_3 \tilde{k}_4 \over 2} 
(\tilde{k}_1 + \tilde{k}_2)_{\n}
\, J_2 (k_1, k_2 )
\Bigr]
\\ \nonumber \cr 
&&
\  \ + \ \ {\rm permutations}\, 
\Bigr\}
 \ .   
\eeqa
}
Notice the appearance of the function $J_3 (k_1, k_2 , k_3)$ 
in the previous expression \eqref{S1-4pt-piece}. 

We now compute the contribution to the four-point function 
derived from the term \eqref{new_new_term}. 
After some straightforward calculations, one gets:
{\small
\beqa
\label{S2-4pt-piece}
&&
\left. 
\Gamma_{\mu \n \r \s}(k_1, k_2,k_3,k_4)
\right|_{S^{
(2)}_{\rm eff}}
\, = \,  \, \, 
\!\!\!\! 
(2\pi)^4\,\delta^{(4)}(k_{1}+ \cdots + k_{4}) 
\, 
\Bigl\{ 
\,2 T(k_4) \, \tilde{k}^{4}_{\s} \cdot
\\ \nonumber \cr 
&&  
\!\!\!\! \cdot
\Bigl[
- Q^{1}_{\mu , \rho \n}
{k_2 \tilde{k}_3 \over  2} J_2 (k_2 , k_3) 
\cos {k_1 \tilde{k}_4 \over  2}
 - Q^{2}_{\n , \r \mu}
{k_1 \tilde{k}_3 \over  2} J_2 (k_1 , k_3) 
\cos {k_2 \tilde{k}_4 \over  2} 
- 
Q^{3}_{\r , \mu \n}
{k_2 \tilde{k}_1 \over  2} J_2 (k_1 , k_2) 
\cos {k_3 \tilde{k}_4 \over  2} 
\Bigr]
\\ \nonumber \cr 
&&
+ {1\over 2} \, P^{12}_{\n \mu} \, \tilde{k}^{4}_{\r}
\, J_2 (k_3 , k_4 ) \, \cos {k_1 \tilde{k}_2 \over 2} \, +
\,  
{1\over 2} \, P^{13}_{\r \mu} \, \tilde{k}^{4}_{\n}
\, J_2 (k_2 , k_4 ) \, \cos {k_1 \tilde{k}_3 \over 2} \, +
\,  
{1\over 2} \, P^{23}_{\r \n} \, \tilde{k}^{4}_{\mu}
\, J_2 (k_1 , k_4 ) \, \cos {k_2 \tilde{k}_3 \over 2} \Bigr]
\\ \nonumber \cr 
&&
-{1\over 2} \, \Bigl[
T(k_1 + k_2) J_2 (k_1 , k_2 ) \cos {k_3 \tilde{k_4} \over 2}
(\tilde{k}_1 + \tilde{k}_2)_{\m} 
(\tilde{k}_1 + \tilde{k}_2)_{\n}
P^{34}_{\s \r} + 
\\ \nonumber \cr 
&&
\qquad \! 
T(k_1 + k_3) J_2 (k_1 , k_3 ) \cos {k_2 \tilde{k_4} \over 2}
(\tilde{k}_1 + \tilde{k}_3)_{\mu} 
(\tilde{k}_1 + \tilde{k}_3)_{\r}
P^{24}_{\s \n} +
\\ \nonumber \cr 
&&\, \, 
\qquad \!
T(k_1 + k_4) J_2 (k_1 , k_4 ) \cos {k_1 \tilde{k_4} \over 2}
(\tilde{k}_1 + \tilde{k}_4)_{\m} 
(\tilde{k}_1 + \tilde{k}_4)_{\s}
P^{32}_{\n \r}
\Bigr] \ + \ {\rm permutations} \ \Bigl\}
\ .
\eeqa
}
\section{A lightning review of the background field method}
We will now apply the background field method to the study of 
correlators in a generic noncommutative theory 
with gauge group $U(N)$.
Our analysis follows closely the approach of 
\cite{KT}, to which we refer the reader for details 
of the application of the background field method
to noncommutative theories.

We will decompose
the gauge field $A_\m$ as $A_\m = B_\m + Q_\m$, where 
 $B_\m$ is  a slowly varying background field, and
$Q_\m$ represents the high-energy fluctuations.
After functionally integrating over the high-frequency fields,
we are left with an effective action for the 
background fields $B_\m$. 
Importantly, this effective action is invariant with respect 
to gauge transformations of the background field, 
$B \to \Omega (B + \partial )\Omega^{\dagger}$, 
where $\Omega$ is an element of the noncommutative  $U(N)$, 
$\Omega := e_{\star}^{\a_A T^A}$.

The action functional which describes the dynamics of  
a spin-$j$ noncommutative field 
$\phi_{m,a}$ in the representation  
{\bf r} of the gauge group in the background of $B_\mu$ has the general
form \cite{KT,Peskin}
\beqa 
\label{gen_f}
S [ \phi ] &=&  
-\int\!  d^{4} x \  \phi_{m,a} \star  
\left ( - D^2 (B)\delta_{mn}\delta^{ab} + 2i (F_{\mu \nu}^B)^{ab}  
\hf J^{\mu \nu}_{mn}   
\right)\star \phi_{n,b} 
\cr 
&\equiv&-\int\!  d^{4} x \   
\phi_{m,a} \star [ \Delta_{j,{\bf r}}]_{mn}^{ab}\star \phi_{n,b} 
\ \ .  
\eeqa 
Here $a,b$ are indices of the  representation {\bf r} of 
noncommutative $U(N)$,  
$F^{ab}\equiv \sum_{A=1}^{N^2} F^A t^A_{ab}$, and $m,n$ 
are spin indices and  
$J^{\mu \nu}_{mn}$ are  
the generators of the euclidean Lorentz group appropriate  
for the spin of the field $\phi$, 
i.e.~$J \, =\,  0$ for spin $0$ fields, 
$J^{\mu \nu}_{\rho \sigma}\, =\, 
i\, (\delta^\mu_\rho \, \delta^\nu_\sigma \, - \,  
\delta^\nu_\rho\, \delta^\mu_\sigma)$ for vectors, and 
$[J^{\mu \nu}]_{\alpha}^{\ \beta} \, = \, 
\ihf \, [\sigma^{\mu \nu}]_{\alpha}^{\ \beta}$
for Weyl fermions.

We 
consider a generic supersymmetric theory 
with {\it adjoint} chiral multiplets,
therefore we  only need to know the action of 
$\Delta_{j, {\bf r}}$   on  
adjoint  fields. 
In this case, it easy to see from  \eqref{gen_f} 
that $\Delta_{j, {\bf \scriptscriptstyle G}}\star \phi$ gives
\begin{equation}
\label{deladj}
\Delta_{j, {\bf \scriptscriptstyle G}}\star \phi
=-\partial^2 \phi -  \left[ (\partial_\mu B_\mu ) , \phi \right]_\star -
2 \left[ B_\mu \partial_\mu , \phi \right]_{\star}-
\left[ B_\mu ,  \left[ B_\mu , \phi \right]_\star  \right]_{\star}
+ 2i \left( \frac{1}{2} J^{\mu \nu}
\left[ F_{\mu \nu}^B , \phi \right]_{\star}
\right)
\ .
\end{equation}
The  one-loop expression for the 
effective action  reads \cite{KT}
\begin{equation}
\label{action_start}
S_{\rm eff} [B] = -\frac{1}{2g^2} \int\!  d^{4} x \ 
\Tr \, 
F_{\mu \nu}^B\star  F_{\mu \nu}^B - \sum_{j,{\bf r}} \alpha_{j} \log \, 
{\rm det}_{\star} \Delta_{j, {\bf r}} 
\ .
\end{equation}
The sum in \eqref{action_start} is extended to all fields in the theory,
including ghosts and gauge fields.
$\alpha_{j}$ is equal to $+1$ ($-1$) for ghost (scalar) fields
and to $+1/2$ ($-1/2$) for
Weyl fermions (gauge fields).
Finally, the functional star-determinants are computed by 
\beqa
\label{deflogk}
\log \, {\rm det}_{\star} \Delta_{j, {\bf r}} 
&\equiv&
\log \, {\rm det}_{\star} \left[-\partial^2 + {\cal K}(B)_{j, {\bf r}}\right]
\\ \nonumber
&= &\log \, {\rm det}_{\star} (-\partial^2) + 
{\rm tr}_{\star} \log \, 
\left[1+(-\partial^2)^{-1}{\cal K}(B)_{j, {\bf r}}\right] 
\ .
\eeqa
The first term on the second line of \eqref{deflogk} 
contributes only to the vacuum loops
and will be dropped in the following.
The second term on the last line of \eqref{deflogk} 
has a full  expansion in terms of Feynman diagrams, 
and on this term we concentrate our attention. 
\begin{figure}[ht]
\centering
\includegraphics{3pt_n.epsi}
\vspace{3mm}
\includegraphics{3pt_b.epsi}
\vspace{3mm}
\includegraphics{4pt_n.epsi}
\vspace{3mm}
\includegraphics{4pt_b.epsi}
\caption{\it Feynman rules for $U(N)$.
The wavy lines represent the background fields, the dotted lines
the high-virtuality fields. }
\label{feyn-rules}
\end{figure}

\subsection{Feynman Rules}
We follow the conventions of \cite{HKT}, whose Feynman rules 
we will use. We  only need to compute one  additional
interaction vertex, which 
originates  from the commutator term in  
the field strength $F_{\m \n}^{B}$ appearing in the last term of
\eqref{deladj}. 
Using \eqref{field-comm} and  \eqref{star-prod} 
this term can be written as: 
\begin{multline}
2i \Tr \int\!  d^{4} x \; \overline{\phi} J^{\mu \nu} \star
[[B_{\mu},B_{\nu}]_{\star},\phi ]_{\star} \, = \, \int\!\!  \frac{d^{4}
p'}{(2 \pi)^4} \frac{d^{4} q_{1}}{(2 \pi)^4} \frac{d^{4} q_{2}}{(2
\pi)^4} \frac{d^{4} p}{(2 \pi)^4} (2 \pi)^4 \delta^{(4)}
\Bigl(p'+q_{1}+q_{2} +p \Bigr) 
\\
\shoveright{ \cdot \,  \bar{\phi}^{A} (p' )
    B_{\mu}^{B}(q) B_{\nu}^{C}(q_{2})\phi^{D} (p )
    \Bigl[ -i J^{\mu \nu} \Bigl(- d^{ B C E} 
\sin \Bigl( \frac{q_{1} \tilde{q}_{2}}{2}
    \Bigr) + f^{B C E} \cos \Bigl(\frac{q_{1} \tilde{q}_{2}}{2} 
    \Bigr) \Bigr) } \nonumber \\
    \cdot \, \Bigl(f^{E D A} 
\cos \Bigl(\frac{p \tilde{p}'}{2} 
    \Bigr) + d^{ E D A} \sin\Bigl( \frac{p \tilde{p}'}{2}
 \Bigr)
    \Bigr) \Bigr]
\ .
\end{multline}
The complete set of Feynman rules 
is shown in figure \ref{feyn-rules}.

\noindent
We denote by  a triangle and  
a star  the so-called 
``$J$-vertices'', second and fourth line
in the Feynman rules of figure \ref{feyn-rules} 
respectively, which are specific to the 
background field method.
\section{Perturbative calculations in the microscopic theory}
We now move on to the background field method 
computation of Green's functions in the microscopic theory. 
For convenience, we present the three-point function 
and four-point function calculations separately. 
\subsection{The three-point function of background fields}
We start with the calculation of 
the three-point function  of background gauge fields
$\Gamma_{\m \n \r}^{ABC} (k_1, k_2, k_3)$
defined in \eqref{tpf}.
\begin{figure}[ht]
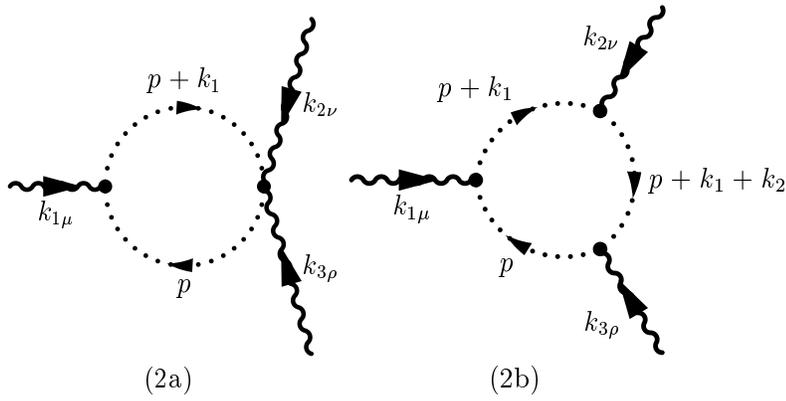

\centering
\includegraphics[width=4.4cm]{2v_4n_3n.epsi}
\includegraphics[width=5.8cm]{3v_3n_0b.epsi}
\caption{\it Feynman diagrams with no $J$-vertices.}
\label{feyn-none}
\end{figure}
To this end, we will need to expand the logarithm 
in  \eqref{action_start} 
up to three powers of the background field. 
The resulting Feynman diagrams 
are shown in figures \ref{feyn-none}-\ref{feyn-three} 
(where we do not draw permutations of the diagrams).

The Feynman diagrams can be conveniently classified
according to the number of $J$-vertices they contain.  
Diagrams with no
$J$-vertices,  represented in 
figure \ref{feyn-none},  
give a vanishing  contribution to the correlator. 
This is because each of these diagrams
gets a factor of $\Tr \, \uno_j \equiv d(j)$ 
from the trace over spin indices, where $d(j)$ is the number 
of spin components of the field, $d(j)=1$ for scalars, 
$2$ for Weyl fermions and $4$ for gauge fields, respectively.
We focus only on supersymmetric theories, where
the cancellation between fermionic and bosonic degrees 
of freedom implies that
\begin{equation}
\sum_{j} \alpha_j \, d(j) = 0
\ . 
\end{equation}
Therefore, 
each diagram which  no $J$-vertices vanishes separately 
when it is summed over all the fields in the theory.
\begin{figure}[ht]
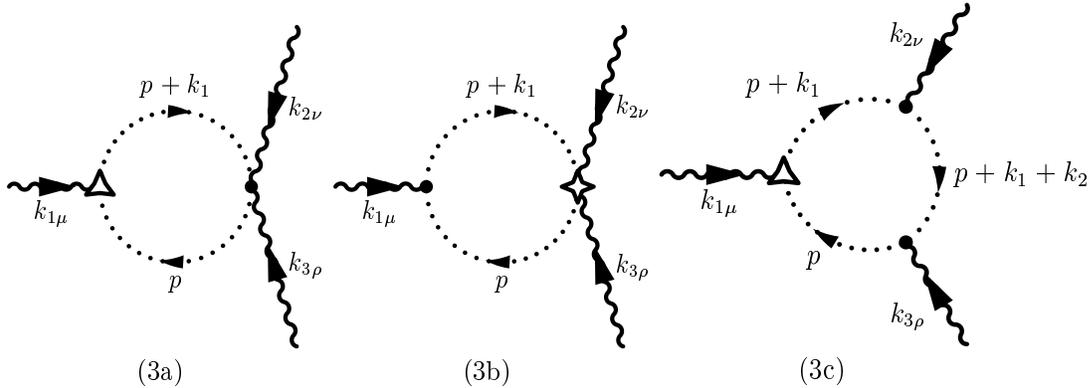

\centering
\includegraphics[width=4.2cm]{2v_4n_3b.epsi}
\includegraphics[width=4.2cm]{2v_4b_3n.epsi}
\includegraphics[width=5.7cm]{3v_2n_1b.epsi}
\caption{\it Feynman diagrams with a single insertion of 
$J$-vertices (denoted by a triangle and a star).}
\label{feyn-one}
\end{figure}
Similarly, 
diagrams with exactly one insertion of the $J$-vertices 
(figure \ref{feyn-one}) vanish, since  
the trace over spin indices gives 
${\rm Tr} \, J^{\mu \nu}=0$.

With these simplifications, we are left  with  
the diagrams of figures  \ref{feyn-two} and 
\ref{feyn-three}, which we now compute. 
We will calculate these diagrams in a low-energy 
approximation where the 
background fields have a much smaller momentum than the 
cut-off for the fluctuating fields, so that 
$k_{i} k_{j} \to 0$,  while we keep $k_{i} \tilde{k}_{j}$ finite
\cite{Zanon,Santambrogio:2000rs,Pernici:2000va}. 
This low-energy approximation has the great 
advantage that all of the integrals over 
the loop momentum
can be performed explicitly
\cite{Liu:2000ad,Zanon} 
(see also the discussion after \eqref{ellel}).

We first consider diagrams with two $J$-vertices, 
represented in figure \ref{feyn-two}.
\begin{figure}[ht]
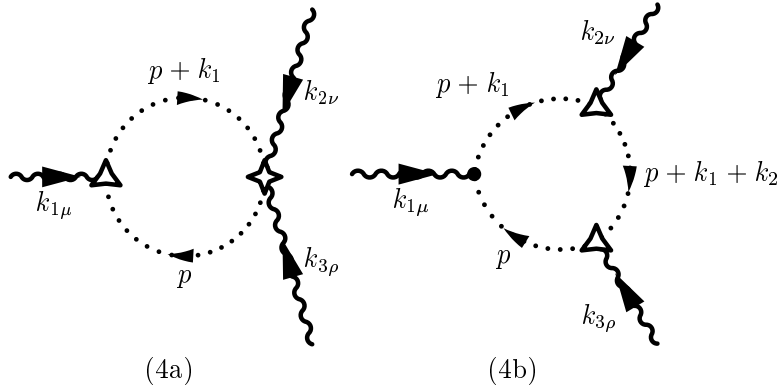

\centering
\includegraphics[width=4.4cm]{2v_4b_3b.epsi}
\includegraphics[width=5.7cm]{3v_1n_2b.epsi}
\caption{\it Feynman diagrams with two $J$-vertices.}
\label{feyn-two}
\end{figure}
The contribution to the correlator
$\Gamma_{\m \n \r}^{ABC}(k_1 , k_2 , k_3)$ 
from diagram (\ref{feyn-two}a) is:
\beqa
\nonumber
&& -2i 
\Bigl(
\sum_{j} \a_j\,\Tr \,  
(J^{\n \r} J^{\alpha \mu})_j 
\Bigr)
\, 
\, (2\pi)^4\delta^{(4)}(k_1 +k_2 +k_3)
\int\!\! \frac{d^4 p}{(2\pi)^4}
\, 
k_{1 {\alpha}} \frac{\delta^{DF}}{p^{2}}
\frac{\delta^{EG}}{(p+k_1)^2} 
\\ \nonumber
&& 
\Bigl[-d^{EAD}\sin\Bigl(\frac{k_1 \tilde{p}}{2}\Bigr) +
f^{EAD} \cos\Bigl(\frac{k_1 \tilde{p}}{2}\Bigr) \Bigr]
\Bigl[-d^{CBH}\sin\Bigl(\frac{k_2 \tilde{k}_3}{2}\Bigr) +
f^{CBH} \cos\Bigl(\frac{k_2 \tilde{k}_3}{2}\Bigr) \Bigr]
\\ 
\label{bbbb}
&&
\Bigl[d^{HGF}\sin\Bigl(\frac{(p+k_1)\tilde{p}}{2} \Bigr) +
f^{HGF} \cos\Bigl(\frac{(p+k_1) \tilde{p}}{2}(p+k_1) \tilde{p}\Bigr)
\Bigr]
\ , 
\eeqa
where the sum is over all the fields in the theory.
We can simplify the products of $d$'s  and $f$'s 
in \eqref{bbbb}
by using  the relations derived in 
\mbox{(2.8)}--\mbox{(2.11)} of 
\cite{Bonora}. In addition,  
the product of $J$'s can be rewritten 
using 
\begin{equation}
\Tr (J^{\mu \rho} J^{ \nu \lambda})_{j}\  =\  
C(j) \, (\delta^{\mu \nu} \delta^{\rho \lambda} -
\delta^{\mu \lambda} \delta^{\nu \rho})
\ ,
\end{equation}
where
\begin{equation}
C(j) \equiv \quad 0 \quad {\rm for\, scalars,}\qquad \quad
\frac{1}{2}\quad {\rm for\, Weyl\,\,  
fermions,} \qquad 2 \quad{\rm
for\, vectors}.
\end{equation}
The remaining integrals can then be evaluated by first 
writing the sines and cosines 
in terms of exponentials,  and then using
\begin{equation}
\label{tbmo}
\int\!\!  {\frac{d^4 p}{(2\pi)^4}}\, 
{\frac{e^{i p\tilde{k}}}{p^2 (p+k)^2}}  = 
{\frac{2}{(4\pi)^2}} \int_{0}^{1}\!\! dx \
K_{0} (\sqrt{A} |k| |\tilde{k}|)
\, \equiv\,  T(k)
\ \ ,
\end{equation}
where $A= x(1-x) $ and 
the function $T$ is the same as in \eqref{ti}.
In this way, 
the contribution to the three-point function from 
diagram (\ref{feyn-two}a)  (and its  permutations)
becomes:  
\beqa 
\label{diag_d_answer} 
[\Gamma_{\mu \nu \rho}^{A B C}(k_1,
k_2, k_3)]_{\rm 4 a} 
&=&  4i 
\, 
\Bigl(
\sum_{j} \a_j \, C ( j ) \Bigr)
\,  \sqrt{2N} 
\, 
(2 \pi)^4 \delta^{(4)}(k_1 +k_2 +k_3) 
\, 
\sin \Bigl( \frac{k_{2} \tilde{k}_3}{2} 
 \Bigr)
\\ \nonumber
\cr
&& 
\!\!\!\!\! 
\Bigl[ 
T(k_1)\, Q^1_{\m , \r \n}\, 
\delta^{A0}  \delta^{BC} 
+ T(k_2)
\,  Q^2_{\n , \m \r}\, 
\delta^{B0} \delta^{AC}  
+ T(k_3) \, 
 Q^3_{\r , \n \m} \delta^{C0} \delta^{AB} \, 
\Bigr]
\  .
\eeqa
%
%
%
%
%
Diagram (\ref{feyn-two}b) 
contributes to the correlator
$\Gamma_{\m \n \r}^{ABC}(k_1 , k_2 , k_3)$ 
as:
\begin{equation}
\label{diag-g-start}
\begin{aligned}
&4 i \, 
\Bigl(
\sum_{j} \a_j\, \Tr
\,  (J^{\a \nu} J^{\b \r})_j \Bigr)
\, 
 \, 
(2 \pi)^4 \delta^{(4)}(k_1+k_2+k_3)
\, 
\\
&
\int\!\! 
\frac{d^4 p }{(2 \pi)^4} \, k_{2 \alpha} k_{3 \beta} (2p +k_1)_\mu
\, \frac{\delta^{EF}}{p^2}\frac{\delta^{GH}}{(p+k_1)^2}
\frac{\delta^{IJ}}{(p+k_1+k_2)^2}
\\
& 
\Bigl[-d^{EAJ} \sin\Bigl( \frac{k_1 \tilde{p}}{2}\Bigr) +
f^{EAJ} \cos\Bigl(\frac{k_1 \tilde{p}}{2}\Bigr)\Bigr] 
\Bigl[ -d^{GBF}
\sin\Bigl(\frac{k_2 (\tilde{p} + \tilde{k}_1) }{2} \Bigr) + f^{GBF}
\cos\Bigl(\frac{k_2 (\tilde{p}+\tilde{k}_1)}{2}\Bigr)\Bigr]
\\
& 
\Bigl[
-d^{ICH} \sin\Bigl(\frac{k_3 (\tilde{p}+\tilde{k}_1+\tilde{k}_2)}{2}
\Bigr) + f^{ICH} \cos\Bigl(\frac{k_2 (\tilde{p}+\tilde{k}_1 +
\tilde{k}_2)}{2}
\Bigr)\Bigr]
\ .
\end{aligned}
\end{equation}
First, we rewrite   
sines  and cosines in terms of exponentials 
in the same way as for diagram 
(\ref{feyn-two}a). 
We will then need to evaluate  integrals of the form
\begin{equation}
\label{povl}
\cL  \left( \sigma , \beta, \gamma \right) =\int\!\!  {\frac{d^4
p}{(2\pi)^4}}{\frac{e^{i p\tilde{\sigma}}}{p^2 (p+\beta)^2 (p+
\beta + \gamma)^2}}
 \ . 
\end{equation}
In the background field method, we integrate out highly
fluctuating momenta;    here, it will be extremely convenient to 
integrate momenta above an infrared scale $\mu$. 
Effectively, this amounts to   introducing
a small mass  term $\mu^2$ in   each propagator, 
so that   \eqref{povl} is turned into 
\begin{equation}
L \left( \sigma , \beta, \gamma \right) =\int\!\!  {\frac{d^4
p}{(2\pi)^4}}{\frac{e^{i p\tilde{\sigma}}}{(p^2+\mu^2)
[(p+\beta)^2+\mu^2] [(p+ \beta + \gamma)^2+\mu^2]}}
\ . 
\end{equation}
Introducing Schwinger parameters, we can recast this integral as
\beqa
L \left( \sigma , \beta, \gamma \right) &=&
\int\!\!  {\frac{d^4 p}{(2\pi)^4}} \int_{0}^{\infty}\!\!
d\alpha_{i} \
{e^{i p\tilde{\sigma}}} e^{
-\alpha_{1}(p^2+\mu^2) - \alpha_{2}((p+\beta)^2+\mu^2)
   -\alpha_{3}((p+ \beta + \gamma)^2+\mu^2)
 }
\\ \nonumber
&=&
\int\!\!  {\frac{d^4 p}{(2\pi)^4}} \int_{0}^{\infty}
\!\! d\alpha_{i}
\, 
\exp ( -\alpha \mu^2 )
   \exp
\Bigl[ i \Bigl( l - \frac{1}{\alpha} (\beta \alpha_{2} +
\alpha_{3}(\beta+\gamma))\Bigr) \tilde{\sigma}\Bigr]
\\ \nonumber
&&
\exp\Bigl[
   -\alpha l^2 +\frac{1}{\alpha}\Bigl(
   -\alpha_{1}\left(\alpha_{2}+\alpha_{3}\right)\beta^2
      -\alpha_{3}\left(\alpha_{1} +\alpha_{2}\right)\gamma^2
      -2\alpha_{1}\alpha_{3} \beta \gamma
      \Bigr) \Bigr]
\ , 
\nonumber
\eeqa 
where 
$\alpha=\alpha_{1}+\alpha_{2}+\alpha_{3}$ 
and
$l=p+
\frac{1}{\alpha}
\left[\beta \alpha_{2} + \alpha_{3}
\left(\beta + \gamma \right) \right]$. 
Following \cite{Zanon}, we change variables to 
$\xi_{i}= \alpha_{i} / \alpha$  and add
a new integration variable  $\lambda$ with a delta function, 
$\delta \left( \lambda - \sum_{i} \alpha_{i}\right)$.
After performing the loop momentum integration, we obtain: 
\beqa
\label{ellel}
L\left( \sigma , \beta, \gamma \right) =
\int_{0}^{\infty}\frac{d\lambda}{\left(4 \pi\right)^2} \, 
e^{- \tilde{\sigma}^2  / (4 \lambda) - \lambda \mu^2
}
\int_{0}^{1} \!\! d\xi_{1} \int_{0}^{1-\xi_{1}} 
\!\!
d \xi_{3} \ \
e^{-i\left[ \left(1-\xi_{1} \right) \beta + \gamma \xi_{3} \right]
\tilde{\sigma}}
\nonumber \\
\cdot \ e^{-\lambda \left[
\xi_{1} \left(1-\xi_{1} \right) \beta^2 +
\xi_{3}\left(1-\xi_{3}\right)\gamma^2
-2\xi_{1} \xi_{3} \beta \gamma
\right]}
\ . 
\eeqa
In the low-energy approximation
we are considering, where 
$k_{i} k_{j} \to 0$ while
$k_{i} \tilde{k}_{j}$ is kept finite, 
the integration becomes feasible 
\cite{Liu:2000ad,Zanon},
and the results for the required cases are:
\beqa
L\left( \sigma , \sigma, \gamma \right) & = & 
M \left(\m |\tilde{\sigma}|\right)
\left(
\frac{1- e^{-i\gamma\tilde{\sigma}}}{\left(\gamma \tilde{\sigma} \right)^{2}}
- \frac{i}{\gamma \tilde{\sigma}} \right)
\ , 
\\ \nonumber  \\
L\left( \sigma , \beta, \sigma \right) &=& 
M \left(\m |\tilde{\sigma}|\right)
\left( i\frac{e^{-i\beta\tilde{\sigma}}}{\beta \tilde{\sigma}} +
\frac{e^{i \beta \tilde{\sigma}} -1}{\left(\beta \tilde{\sigma}\right)^2}
 \right)
\ , 
\eeqa
and the case where
$\sigma \ne \beta \ne \gamma$ but  $\sigma + \beta +\gamma = 0$:
\begin{equation}
\left[\, L \left(\sigma , \beta, \gamma \right)\, 
\right]_{\s + \b + \gamma = 0} 
\, = \, 
M \left(\m |\tilde{\sigma}|\right)
\left( \frac{i}{\gamma \tilde{\sigma}} + 
\frac{e^{-i\beta \tilde{\sigma}} - 1}{\left(\gamma \tilde{\sigma}\right) 
\left(\beta \tilde{\sigma}\right)}\right)
\ ,
\end{equation}
where we have defined
\begin{equation}
M \left(\m |\tilde{\sigma}|\right)
:=\int_{0}^{\infty}\!\!
\frac{d\lambda}{\left(4 \pi\right)^2}\ 
e^{- \tilde{\sigma}^2  / (4 \lambda) - \lambda \mu^2
}
\ . 
\end{equation}
We also need a  variant of the $L$ integral with an extra power of
$p_\m$ in the numerator. We calculate this by noting that
\beq 
\int\!\! 
{\frac{d^4 p}{(2\pi)^4}}\, {\frac{p_\mu \, 
e^{i p\tilde{\sigma}}}{p^2
(p+\beta)^2 (p+ \beta + \gamma)^2}} \, =\, 
-i
\frac{d}{d\tilde{\sigma}^{\mu}}\int\!\!  {\frac{d^4
p}{(2\pi)^4}}\, {\frac{e^{i p\tilde{\sigma}}}{p^2 (p+\beta)^2 (p+
\beta + \gamma)^2}} \ . 
\eeq 
After some algebra, the contributions of  diagram 
(\ref{feyn-two}b)
 and its
permutations  to the three-point function becomes, 
in the low-energy approximation we are considering: 
\beqa
\label{diag_g_answer_old} 
[\Gamma_{\m \n \r}^{ABC}(k_1 , k_2 , k_3 )]_{\rm 4 b}
&=& 
8i \sqrt{2N} 
\, \Bigl(
\sum_{j} \a_j\, C ( j )  \Bigr)
\frac{\sin (  k_{2} \tilde{k}_3 / 2 )}{
k_{2}\tilde{k}_3}(2 \pi)^4 \delta^{(4)}(k_1 +k_2 +k_3)
\\
&& 
\!\!\!\!\!\!
\biggl\{
P_{\r \n}^{23}\, 
\left(\dot{M}_{\mu} (\m |\tilde{k}_{1}| ) \ \delta^{A0} \delta^{BC} 
+
\dot{M}_{\mu}(\m |\tilde{k}_{2}| )\  \delta^{B0} \delta^{AC}  +
\dot{M}_{\mu} (\m |\tilde{k}_{3}| ) \ \delta^{C0} \delta^{AB}  \right)
\nonumber \\
&&
\!\!\!\!\!\!
\ \ P_{\r \m}^{13}\, 
\left(\dot{M}_{\nu} (\m |\tilde{k}_{1}| )\
\delta^{A0} \delta^{BC} + \dot{M}_{\nu}  (\m |\tilde{k}_{2}| ) \
\delta^{B0}\delta^{AC} + \dot{M}_{\nu}  (\m |\tilde{k}_{3}| )\
\delta^{C0} \delta^{AB} \right)
\nonumber \\
&&
\!\!\!\!\!\!
\ \ P_{\n \m}^{12}
\left(\dot{M}_{\rho} (\m |\tilde{k}_{1}| )\
 \delta^{A0}\delta^{BC} +\dot{M}_{\rho} ( \m |\tilde{k}_{2}| )\
\delta^{B0}\delta^{AC} +\dot{M}_{\rho} (\m |\tilde{k}_{3}| )\
\delta^{C0} \delta^{AB} \right)
\biggr\}
\ , 
\nonumber
\eeqa
where
$
\dot{M}_\mu(z):=(d M / d z^\m ) (z).
$
Since 
$
\dot{M}_\mu (z ) \ =\
(z_\mu /  {2 }) \,  S (z )
$, 
where 
\begin{equation}
\label{esse}
 S(z )\ =\ \frac{2}{(4 \pi)^2} \, 
K_0(z )
\ , 
\end{equation}
we can finally recast \eqref{diag_g_answer_old} as 
\beqa
\label{diag_g_answer} 
[\Gamma_{\m \n \r}^{ABC}(k_1 , k_2 , k_3 )]_{\rm 4 b}
\!\!\!\!\!
&&=\ 
4 i\sqrt{2N} 
\, \Bigl(
\sum_{j} \a_j\, C ( j )  \Bigr)
\frac{\sin (  k_{2} \tilde{k}_3 / 2 )}{
k_{2}\tilde{k}_3}(2 \pi)^4 \delta^{(4)}(k_1 +k_2 +k_3)
\\
&& \!\!\!\!\!\!\!\!\!\!\!\!
\biggl\{
P_{\r \n}^{23}\, 
\left(\tilde{k}_{1\m}
S (\m |\tilde{k}_{1}| ) \ \delta^{A0} \delta^{BC} 
+
\tilde{k}_{2\m } S(\m |\tilde{k}_{2}| )\  \delta^{B0} \delta^{AC}  +
\tilde{k}_{3\m}S(\m |\tilde{k}_{3}| ) \ \delta^{C0} \delta^{AB}  \right)
\nonumber \\
&&\!\!\!\!\!\!\!\!\!\!\!\!
\ \ P_{\r \m}^{13}\, 
\left(\tilde{k}_{1\n } S (\m |\tilde{k}_{1}| )\
\delta^{A0} \delta^{BC} + 
\tilde{k}_{2\n} S (\m |\tilde{k}_{2}| ) \
\delta^{B0}\delta^{AC} + 
\tilde{k}_{3\n} S(\m |\tilde{k}_{3}| )\
\delta^{C0} \delta^{AB} \right)
\nonumber \\
&&\!\!\!\!\!\!\!\!\!\!\!\!
\ \ P_{\n \m}^{12}
\left(
\tilde{k}_{1\r} S(\m |\tilde{k}_{1}| )\
 \delta^{A0}\delta^{BC} +
\tilde{k}_{2\r} S ( \m|\tilde{k}_{2}| )\
\delta^{B0}\delta^{AC} +\tilde{k}_{3\r} S (\m |\tilde{k}_{3}| )\
\delta^{C0} \delta^{AB} \right)
\biggr\}
\ .
\nonumber
\eeqa
The last diagram to compute is shown in figure 
\ref{feyn-three}.
\begin{figure}[ht]
\centering
\includegraphics[width=5.7cm]{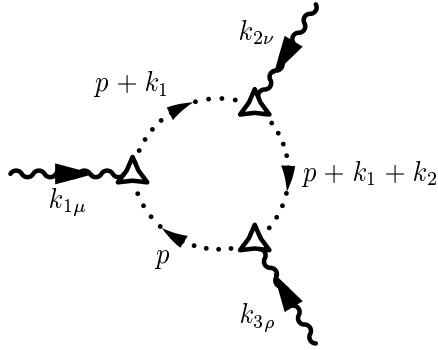}
\caption{\it This diagram contains three insertions of the 
triangle $J$-vertex.}
\label{feyn-three}
\end{figure}
It is easily seen from 
the Feynman rule of the ``triangle'' $J$-vertex 
that  this diagram gives a subleading contribution in 
the low-energy approximation
$k_{i} k_{j} \to 0$ and 
$k_{i} \tilde{k}_{j}$ finite, 
when compared to the diagrams
(\ref{feyn-two}a), (\ref{feyn-two}b)
 computed so far, hence  we will discard its contribution. 
Summarising, the full three-point function is 
obtained by adding up the results  
\eqref{diag_d_answer}   and \eqref{diag_g_answer}.

\subsection{The four-point function of background fields}
In this section we present  the calculation of 
the four-point function of background fields
in the microscopic theory. 
The computation  proceeds in much the same way as that of 
the three-point function presented in the previous section. 
For simplicity, we will limit ourselves to the case of gauge group 
$U(1)$. 

As in the three-point function case, 
only diagrams with at least two insertions of $J$-vertices 
give a nonvanishing contribution 
(in the supersymmetric theories we are interested in). 
\begin{figure}[ht]
\centering
\includegraphics[width=4.4cm]{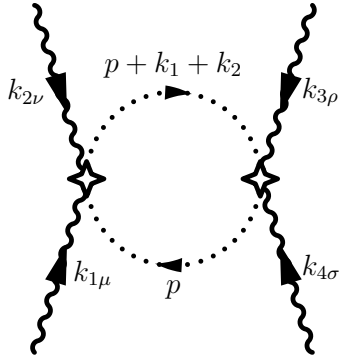}
\label{pic-4pt_1}
\caption{\it This diagram contains two insertions of the star
$J$-vertex.}
\end{figure}
\begin{figure}[ht]
\centering
\includegraphics[width=4.4cm]{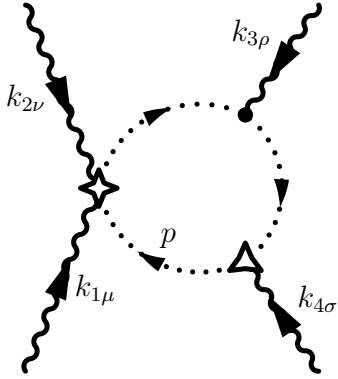}
\label{pic-4pt_2}
\caption{\it Feynman diagram containing a single insertion of the star and of the triangle $J$-vertex.}
\end{figure}
Furthermore, terms in the effective action expressions without
the functions $P$ or $Q$ (defined in \eqref{PQ}) must arise
from diagrams containing  
no powers of external momenta in their vertices. 
The only such candidates are therefore the diagram shown in figure 6 
and its permutations. 
The expression for this diagram is proportional  to:
\begin{equation}
4(\delta^{\mu \sigma}
\delta^{\nu \rho}-\delta^{\mu \rho}\delta^{\nu \sigma})
\, 
\mbox{sin}\Bigl(\frac{1}{2} k_1 \tilde{k}_2\Bigr)
\, 
\mbox{sin}\Bigl(\frac{1}{2} k_3 \tilde{k}_4\Bigr) \, 
T(k_1 +k_2)
\ .
\label{4pt-deltapiece}
\end{equation} 
We now consider the  Feynman diagram in figure 7 
(and its permutations). This diagram gives a contribution 
to the  correlator $\Gamma_{\mu \nu \rho \sigma}(k_1,k_2,k_3, k_4)$ 
which is proportional to:
\begin{align}
\label{4pt-qpiece}
&Q^4_{\sigma, \rho \nu} \mbox{sin}\Bigl(\frac{1}{2} k_2 \tilde{k}_3\Bigr)
 J_2(k_1, k_4) \dot{M}_\mu(\mu|\tilde{k}_4|)
+Q^4_{\sigma, \rho \mu} \mbox{sin}\Bigl(\frac{1}{2} k_1 \tilde{k}_3\Bigr)
 J_2(k_2,k_4) \dot{M}_\nu(\mu|\tilde{k}_4|) \\ \nonumber
+&Q^4_{\sigma, \nu \mu} \mbox{sin}\Bigl(\frac{1}{2} k_1 \tilde{k}_2\Bigr)
 J_2(k_3,k_4) \dot{M}_\rho(\mu|\tilde{k}_4|)
+Q^3_{\rho, \nu \mu} \mbox{sin}\Bigl(\frac{1}{2} k_1 \tilde{k}_2\Bigr)
 J_2(k_3, k_4) \dot{M}_\sigma(\mu|\tilde{k}_4|) \\ \nonumber
+&Q^2_{\nu, \rho \mu} \mbox{sin}\Bigl(\frac{1}{2} k_1 \tilde{k}_3\Bigr)
 J_2(k_2, k_4) \dot{M}_\sigma(\mu|\tilde{k}_4|)
+Q^1_{\mu, \rho \nu} \mbox{sin}\Bigl(\frac{1}{2} k_2 \tilde{k}_3\Bigr)
 J_2(k_1, k_4) \dot{M}_\sigma(\mu|\tilde{k}_4|)
\ .
\end{align}
Finally, the remaining diagrams give rise 
to terms which are proportional to the functions $P^{ij}$
defined in \eqref{PQ}. 
In order to calculate these contributions, 
we need the expressions  for a few new integrals.  
Firstly, we need to consider 
the integral $L(\sigma, \beta, \gamma)$,  
defined in \eqref{ellel}, for the case where 
$\sigma \ne \beta \ne \gamma$ but  $\sigma + \beta +\gamma \ne 0$. 
We find that:
\begin{equation}
\left[\, L \left(\sigma , \beta, \gamma \right)\, 
\right]_{\s + \b + \gamma \ne 0} 
\, = \, 
\frac{M \left(\m |\tilde{\sigma}|\right)}{\gamma \tilde{\sigma}}
\left( \frac{e^{-i\beta \tilde{\sigma}}-1}{\beta \tilde{\sigma}} + 
\frac{e^{-i(\beta +\gamma)\tilde{\sigma}} - 1}{\left(\beta + \gamma\right)\tilde{\sigma}}\right)
\ .
\end{equation}
It is also necessary to calculate several integrals 
containing four insertions of propagators.  
These integrals can be evaluated in a similar way 
to that  used in the the calculation
of the integrals appearing in the three-point function calculation. 
For example, one   needs to evaluate,  for  
$\sigma + \beta +\gamma + \delta = 0$,
\begin{align}
L \left( \sigma , \beta, \gamma,\delta \right) = &\int\!\!  {\frac{d^4
p}{(2\pi)^4}}{\frac{e^{i p\tilde{\sigma}}}{(p^2+\mu^2)
[(p+\beta)^2+\mu^2] [(p+ \beta + \gamma)^2+\mu^2][(p+ \beta + \gamma+
\delta)^2+\mu^2]}} \nonumber \\ \cr
&\stackrel{\!\! k_i k_j \to 0}{\longrightarrow}
\frac{N(\tilde{\sigma})}{\tilde{\sigma}(\gamma+\delta)}
\left[
\frac{1-e^{-i\tilde{\sigma}(\beta+\gamma)}}{i\tilde{\sigma}
(\beta+\gamma)}\left(\frac{1}{\tilde{\sigma}\delta}+
\frac{1}{\tilde{\sigma}\gamma}\right)
-\frac{1}{\tilde{\sigma}\delta} - 
\frac{1-e^{-i\tilde{\sigma}\beta}}{i(\tilde{\sigma}\gamma)
(\tilde{\sigma}\beta)}
\right],
\end{align}
where
\begin{equation}
N(\tilde{\sigma})=\frac{1}{16 \pi^2}\int^{\infty}_0 \! d\lambda 
\, \lambda
\, e^{-\lambda\mu^2 - \frac{\tilde{\sigma}^2}{4 \lambda}}
\ , 
\end{equation}
and in the last line we have used the low-energy approximation 
$k_i k_j \to 0$.
\begin{figure}[ht]
\centering
\includegraphics[width=4.4cm]{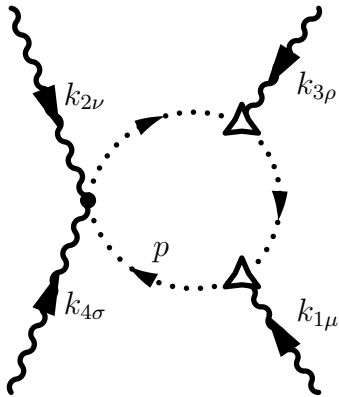}
\label{pic-4pt_3}
\caption{\it This diagram contains two insertions of the triangle $J$-vertex.}
\end{figure}
Using such integrals, one  sees the emergence of 
terms proportional to the $J_2$-
and $J_3$-functions defined in \eqref{jfuncs2} and \eqref{jfuncs3}, 
respectively. 
We skip the details of the calculation, which is rather lengthy 
but,  for example,  the
diagram shown in figure 8 produces a term 
 containing $M(\tilde{k}_4)$ 
and which turns out to be proportional to the expressions
{\small
\begin{equation}
\frac{2}{k_1 \tilde{k}_4}\left[
\frac{\mbox{cos}(\frac{1}{2}(k_1 \tilde{k}_4 -k_2 \tilde{k}_3))-\mbox{cos}(\frac{1}{2}(k_1 \tilde{k}_3 -k_2 \tilde{k}_4))}{k_3 \tilde{k}_4}
+\frac{\mbox{cos}(\frac{1}{2}(k_1 \tilde{k}_3 +k_2 \tilde{k}_4))-\mbox{cos}(\frac{1}{2}(k_1 \tilde{k}_3 -k_2 \tilde{k}_4))}{k_2 \tilde{k}_4}
\right] \ .
\label{BBP}
\end{equation}
}
The previous expression \eqref{BBP} 
is precisely equal to $J_3(k_1, k_2, k_3)$ defined in \eqref{jfuncs3}
after imposing $k_4 = -(k_1 + k_2 + k_3)$. 

\section{Comparison to the result from the effective action }
We are now ready to compare our perturbative results
with the expressions for the three- and four-point 
functions obtained from the effective action
$S_{\rm eff}=S^{(1)}_{\rm eff} + S^{(2)}_{\rm eff}$, where 
$S^{(1)}_{\rm eff}$ and $S^{(2)}_{\rm eff}$ are given in 
\eqref{AL-conj} and \eqref{new_new_term}, respectively.

We begin by considering the three-point function 
of background gauge fields. In this case, 
the full perturbative result
is obtained by summing up 
\eqref{diag_d_answer} 
with \eqref{diag_g_answer}.
We elaborate further these expressions by first 
performing  the sum over the spin $j$. 
For definiteness, 
we consider an $\cN=1$ supersymmetric theory with
$  N_{\rm f}$ adjoint chiral superfields, for which 
\beq
\label{ecco}
\Bigl(
\sum_{j} \a_j \, C ( j ) \Bigr)\, =\, 
-{1\over 4} (3 - N_{\rm f}) 
\ . 
\eeq
The case $N_{\rm f}=0 $ corresponds to pure $\cN=1$ 
super Yang-Mills; 
for $N_{\rm f}=1, 3 $ we have the $\cN=2$ and $\N=4$ 
theories, respectively. Notice that, in the latter case, 
the contribution to the three-point function vanishes.
Secondly, we observe that \eqref{diag_g_answer}
was derived in the low-energy approximation 
$k_{i} k_{j} \to 0$,  with $k_{i} \tilde{k}_{j}$ fixed and finite.
We also introduced a small infrared regulating mass $\m$. 
In order to compute the corresponding limit of 
\eqref{diag_d_answer}, we note that 
this amounts to perform the following modification
on the function $T$ of \eqref{tbmo}:
\beqa
\label{TtoS}
T(k) 
\stackrel{\m}{\longrightarrow}
\int\!\!  {\frac{d^4 p}{(2\pi)^4}}
{\frac{e^{i p\tilde{k}}}{(p^2 + \m^2)  [ (p+k)^2 + \m^2]}}  
\, \stackrel{\!\! k^2 \to 0}{\longrightarrow}\, 
{\frac{2}{(4\pi)^2}} 
\, 
K_{0} (\m |\tilde{k}|)
\, \equiv\,  S(\m |\tilde{k}|)
\ \ ,
\eeqa
where the first arrow stands for equality after 
introducing the regulator $\m$ in the expression for $T$, 
and the second means equality in the 
limit $k^2 \to 0$ (at fixed $|k| |\tilde{k}|$).

Taking these observations into account, 
the one-loop perturbative 
expression 
for the three-point function,
 in the low-energy regime 
$k_{i} k_{j} \to 0$ and 
$k_{i} \tilde{k}_{j}$ finite,
is given by:
\beqa  
\label{complete_feyn_answer}
\Gamma^{ABC}_{\mu \nu \rho}(k_1, k_2,k_3)
& = &   
i\sqrt{2N} (N_{\rm f} - 3) 
\, {\sin ( k_{2}
\tilde{k}_{3} / 2 ) \over 
k_2 \tilde{k}_3}
\ (2 \pi)^4\delta^{(4)}(k_1 +k_2 +k_3) 
\\ \nonumber \cr 
&&\Bigl\{
 \delta^{A0} \delta^{BC}
S(\m|\tilde{k}_1|) 
\left[ 
\tilde{k}_{1 \mu} P_{\r \n}^{23} + 
 \tilde{k}_{1 {\nu}} P^{13}_{\r \m}
 +
 \tilde{k}_{1{\rho}}P^{12}_{\n \m} 
+ (k_2 \tilde{k}_3)\, Q^1_{\m, \n \r} 
\right] 
 \\ \nonumber 
&&+   
 \delta^{B0} \delta^{CA} 
S(\m |\tilde{k}_{2}|)
\left[
  \tilde{k}_{2{\mu}}P^{23}_{\r \n}
+ \tilde{k}_{2 \nu} P_{\r \m}^{13} +
 \tilde{k}_{2{\rho}} P^{12}_{\n \m}  + 
 (k_2 \tilde{k}_3)\, Q^2_{\n, \r \m} 
 \right]
\\ \nonumber 
&&+
 \delta^{C0} \delta^{AB} 
S(\m |\tilde{k}_{3}|)
\left[  
\tilde{k}_{3 {\mu}} P^{23}_{\r \n} + 
 \tilde{k}_{3{\nu}} P^{13}_{\r \m}
+ \tilde{k}_{3 \r} P_{\n \m}^{12}  + 
 (k_2 \tilde{k}_3)\, Q^3_{\r, \m \n} 
\right] 
\ .
\eeqa
This perturbative result
\eqref{complete_feyn_answer}  should be contrasted 
with the result \eqref{a_contr}
(with $\cC = 2 \a_0 / N$) 
obtained from the original expression  \eqref{AL-conj}
for the effective action, where, from the results
of \cite{KT,HKT}, it follows that 
\beq
\label{alfazero}
 \a_0 \, = \, -4 \Bigl(
\sum_{j} \a_j C_j \Bigr) N \, = \,
(3-N_{\rm f}) N
\  , 
\eeq 
the sum over $j$ being extended only to fields in the adjoint.
The expressions \eqref{complete_feyn_answer} 
and \eqref{AL-conj} differ in two respect. 
First, \eqref{AL-conj}
contains the function $T$, whereas 
the perturbative result
\eqref{complete_feyn_answer} contains the function $S$. 
This is easily explained by remembering 
\eqref{TtoS}, i.e.~that at low energy $T\to S$.
Second, and more importantly, the perturbative expression
\eqref{complete_feyn_answer} contains, in addition to the terms 
in \eqref{a_contr}, also a contribution proportional to 
\begin{equation}
\label{extrarc}
\delta^{A0} \delta^{BC}\, S(\m|\tilde{k}_1|)  
\, \tilde{k}_{1 \mu} P_{\r \n}^{23}
\, + \, 
\delta^{B0} \delta^{AC}\, S(\m |\tilde{k}_2|)  
\, \tilde{k}_{2 \nu} P_{\r \m}^{13} 
\, + \,  
\delta^{C0} \delta^{AB}\, S(\m |\tilde{k}_3|)  
\, \tilde{k}_{3 \r} P_{\n \m}^{12}
\ .
\end{equation}
This  new contribution does not arise 
from the originally conjectured action 
$S^{(1)}_{\rm eff}$ in 
\eqref{AL-conj}. However, as we have discussed in 
the introduction, this term
\eqref{extrarc} 
is precisely reproduced  by adding 
the contribution $S^{(2)}_{\rm eff}$
in \eqref{new_new_term}   to the  original 
effective action term in \eqref{AL-conj}. 


Finally,  we consider now  
the matching of the four-point function obtained from the effective
action, Eqs.~\eqref{S1-4pt-piece} and \eqref{S2-4pt-piece},  
against the perturbative calculation presented in the previous section.
We will find that the perturbative calculation 
is precisely reproduced by the effective action 
$S_{\rm eff}=S^{(1)}_{\rm eff} + S^{(2)}_{\rm eff}$, where 
$S^{(1)}_{\rm eff}$ and $S^{(2)}_{\rm eff}$ are the expressions 
in \eqref{AL-conj} and \eqref{new_new_term}.%
\footnote{In particular, the four-point function calculation 
discriminates between the terms \eqref{new_new_term} and 
\eqref{new_term}.}

Feynman diagrams in figure 6 (and its permutations) generate
the contribution 
\eqref{4pt-deltapiece}, which precisely 
matches the terms in our expression \eqref{S1-4pt-piece} 
which contains $\delta_{\m\r} \delta_{\n \s}$ and no insertions
of $P$ and $Q$ functions.
Similarly,  the first three terms in  \eqref{4pt-qpiece} 
precisely reproduce the
terms generated by 
$S^{(1)}_{\rm eff}$  
containing both the $T(k_4)$  and  $Q$ functions. 
The remaining terms in \eqref{4pt-qpiece} correspond to  terms 
produced by $S^{(2)}_{\rm eff}$ 
(see \eqref{S2-4pt-piece}),  when we consider the  
low-energy limit 
$\mbox{cos}(\frac{1}{2}k_i \tilde{k}_j) \to 1$ and 
$\mbox{sin}(\frac{1}{2}k_i \tilde{k}_j) 
\to \frac{1}{2} k_i \tilde{k}_j$.
Finally, as anticipated in the previous section, 
combinations of the remaining Feynman
diagrams reproduce 
those terms in $S_{\rm eff}$ that contain the $P$ function.

Summarising, we have a complete agreement between the
low-energy limit of the perturbative calculation 
of three- and four-point functions in the microscopic 
theory,  and the corresponding result obtained from 
the low-energy effective action
$S_{\rm eff}=S^{(1)}_{\rm eff} + S^{(2)}_{\rm eff}$.

\vspace{1cm}
\section*{Acknowledgements}
Particular thanks go to Valya Khoze for 
illuminating discussions and an enjoyable collaboration 
over a long time,  and for 
comments on this paper. 
We would also like to thank Adi Armoni, Chong-Sun Chu
and Sanjaye Ramgoolam 
for  conversations, and an anonymous Referee 
for important comments.
GT would  like to thank the Theory Group of the 
Physics Department, University of Rome ``Tor Vergata''
for hospitality during the last stage of this work.   
The work of JL was supported by a PPARC studentship and a
scholarship from The Ogden Trust. The work of GT was 
supported by  PPARC.

\newpage
\startappendix
\Appendix{notation and conventions}
We consider theories defined by the noncommutativity relation 
\begin{equation}
[x^\mu, x^\nu] \, = \, i \, \theta^{\mu \nu}
\ , 
\end{equation}
where we choose $\th^{\m \n}$ to be purely 
space-space, i.e.~$\th^{0i}=0$.
The Moyal star-product is defined as 
\beq
(\phi \star \psi) (x)\,  := \, 
\phi(x) \, e^{{i\over 2}\theta^{\mu\nu}
\stackrel{\leftarrow}{\partial_\mu}
\stackrel{\rightarrow}{\partial_\nu}}  \, \psi(x) 
\ . 
\label{stardef}
\eeq
Noncommutative field theories can then be 
regarded as ordinary field theories where the 
usual product of fields is replaced 
by the star-product \eqref{stardef}.
The relation 
\begin{equation}
\label{star-prod}
e^{ik_1 x} \star \cdots \star e^{ik_n x}\  =\ 
e^{-\frac{i}{2} \sum \limits_{i<j}(k_i \tilde{k}_j)}
\, 
e^{i(k_1+ \cdots +k_n) x}
\end{equation}
is also repeatedly used in the calculations, where 
we define $\tilde{k}_\mu :=  \theta_{\mu \nu}k_\nu$.

For calculations involving the gauge group 
$U(N)$, we introduce
anti-hermitian generators 
in the fundamental representation 
as $t^A$, 
$A =(0, a)$, where $a = 1, \ldots , N^2-1$ labels the
$SU(N)$ generators, and $t^0 = ( 1 / i \sqrt{2N}) \uno_{N}$. 
Then
\beq 
\Tr (t^A t^B ) \, = \, 
- \frac{\, \, \delta^{AB}}{2}  
\ . 
\eeq
The generators satisfy
\beq 
[t^A, t^B] = f^{ABC} t^C \, \ ,
\qquad 
\{ t^A, t^B \} = - i d^{ABC} t^C
\ .
\eeq
$f^{ABC}$ ($d^{ABC}$) is completely antisymmetric (symmetric) in its indices;
$f^{abc}$, $d^{abc}$ are the same as in $SU(N)$, and
$f^{0bc}= 0$, $d^{0BC} = \sqrt{2/  N}\  \delta^{BC}$, $d^{00a}= 0$,
$d^{000} = \sqrt{2/ N}$.

Given $\phi_1 = \phi_1^A \, t^A$, $\phi_2 =  \phi_2^A \, t^A$, 
it is convenient to re-express $[\phi_1,\phi_2]_\star$ as 
\begin{equation}
\label{field-comm}
[\phi_1,\phi_2]_\star = {1\over 2}
\left(
\{\phi^A_1, \phi^B_2\}_\star \,  f^{ABC}  
-i [\phi^A_1, \phi^B_2]_\star \, d^{ABC}  
\right) \, t^C
\ . 
\end{equation}
Finally, 
we define our euclidean  $\sigma_\mu$  and $\bar{\sigma}_\mu$ matrices as
$\sigma_\mu=(i\sigma^m , \uno_{2\times 2} )$, and
$\bar{\sigma}_\mu=(-i\sigma^m , \uno_{2\times 2} )$, 
where $\sigma^m$ are the three  Pauli matrices.
We also use
$\sigma_{\mu \nu}= 
\hf (\sigma_\mu \bar{\sigma}_\nu-\sigma_\nu \bar{\sigma}_\mu)= 
i \eta^{a}_{\mu \nu}\sigma^a$, and
$\bar{\sigma}_{\mu \nu}=
\hf 
(\bar{\sigma}_\mu \sigma_\nu-\bar{\sigma}_\nu \sigma_\mu)= 
i \bar{\eta}^{a}_{\mu \nu}\sigma^a$, where $\eta^{a}_{\mu \nu}$ and
$\bar{\eta}^{a}_{ \mu \nu}$
are the self-dual and antiself-dual 't Hooft symbols,
respectively  \cite{'tHooft:fv}.

\newpage

\end{document}